\documentclass[aps,pre,reprint,groupedaddress]{revtex4-1}
\usepackage{graphicx}
\usepackage[textsize=tiny]{todonotes}
\usepackage{dcolumn}
\usepackage{amsmath}
\usepackage{amssymb}
\newcommand{\vect}[1]{\ensuremath{\mathbf{#1}}}
\newcommand{\tens}[1]{\ensuremath{\mathbf{#1}}}

\begin{document}

%\preprint{}

\title{Holographic Measurements of Anisotropic Three-Dimensional
  Diffusion of Colloidal Clusters}
% Alternate titles:
% Manifestations of Broken Symmetries in Experimentally Measured
%     Diffusion Tensors for Anisotropic Colloidal Clusters
% Measurements of Symmetry Breaking in Diffusion Tensors of
%    Anisotropic Colloidal Clusters
% (this is probably a bit too "biology" style for PRL)
% Measured Diffusion Tensors of Anisotropic Colloidal Clusters Reveal
%     Broken Symmetries

\author{Jerome Fung}
\affiliation{Department of Physics, Harvard University, Cambridge, 
Massachusetts 02138, USA}

\author{Vinothan N.~Manoharan}
\email{vnm@seas.harvard.edu}

\affiliation{School of Engineering and Applied Sciences, Harvard
  University, Cambridge, Massachusetts 02138, USA}
\affiliation{Department of Physics, Harvard University, Cambridge, 
Massachusetts 02138, USA}

\date{\today}

\begin{abstract}
  We measure all nonzero elements of the three-dimensional (3D)
  diffusion tensor \tens{D} for clusters of colloidal spheres to a
  precision of 1\% or better using digital holographic microscopy.  We
  study both dimers and triangular trimers of spheres, for which no
  analytical calculations of the diffusion tensor exist.  We observe
  anisotropic rotational and translational diffusion arising from the
  asymmetries of the clusters.  In the case of the three-particle
  triangular cluster, we also detect a small but statistically
  significant difference in the rotational diffusion about the two
  in-plane axes. We attribute this difference to weak breaking of
  threefold rotational symmetry due to a small amount of particle
  polydispersity.  Our experimental measurements agree well with
  numerical calculations and show how diffusion constants can be
  measured under conditions relevant to colloidal self-assembly, where
  theoretical and even numerical prediction is difficult.
\end{abstract}

\pacs{82.70.Dd, 36.40.Sx, 42.40.-i, 42.40.Kw}
%\keywords{}

\maketitle

Diffusion plays a critical role in the dynamics, self-assembly, and
rheology of complex fluids. In systems such as colloidal suspensions,
which typically have short-ranged interaction potentials, diffusion
can in fact play a larger role than energy barriers in setting
transition rates~\cite{holmes-cerfon_geometrical_2013}. However, the
diffusion of geometrically anisotropic particles, a common class of
colloidal suspension that can also arise as intermediates in the
self-assembly of spherical particles, can be difficult to
predict. Theoretically determining friction factors for these
particles requires analytically solving Stokes' equation, which is
only possible for highly symmetric particles such as
ellipsoids~\cite{perrin_mouvement_1934} or sphere
dimers~\cite{nir_creeping_1973} in unbounded fluids. Numerical methods
such as bead modeling~\cite{garcia_de_la_torre_improved_2007} or
finite-element methods~\cite{pakdel_mobility_1991, allison_low_1999,
  aragon_precise_2006} require approximating the shape of the
particles or the hydrodynamic interactions. These methods are
difficult to apply when asymmetric particles diffuse near rigid
boundaries or other particles, two situations that are relevant to
colloidal self-assembly and dynamics in general. Thus experimental
measurements of diffusion tensors are crucial.

In particular, precision measurements on single particles rather than
ensembles are necessary. Anisotropic particles show multiple diffusion
timescales that are difficult to resolve through bulk techniques such
as depolarized dynamic light scattering~\cite{hoffmann_3d_2009}. But
there have been few single-particle studies of anisotropic diffusion
in 3D. Video microscopy has been used to measure two-dimensional (2D)
diffusion of colloidal ellipsoids~\cite{han_brownian_2006,
  han_quasi-two-dimensional_2009} and planar sphere
clusters~\cite{anthony_translation-rotation_2008} but the technique
yields limited information about out-of-plane motions
\cite{speidel_three-dimensional_2003, zhang_three-dimensional_2008,
  colin_3d_2012}.  Confocal microscopy can be used to study the 3D
dynamics of geometrically anisotropic particles, but has only been
applied to highly symmetric
particles~\cite{mukhija_translational_2007, hunter_tracking_2011} and
is limited by the time ($\sim 1$ s) needed to acquire a 3D stack. This
can make it challenging to probe timescales comparable to particle
diffusion times or to study rare processes such as the early stages of
self-assembly~\cite{perry_real-space_2013}.

\begin{figure}
\includegraphics{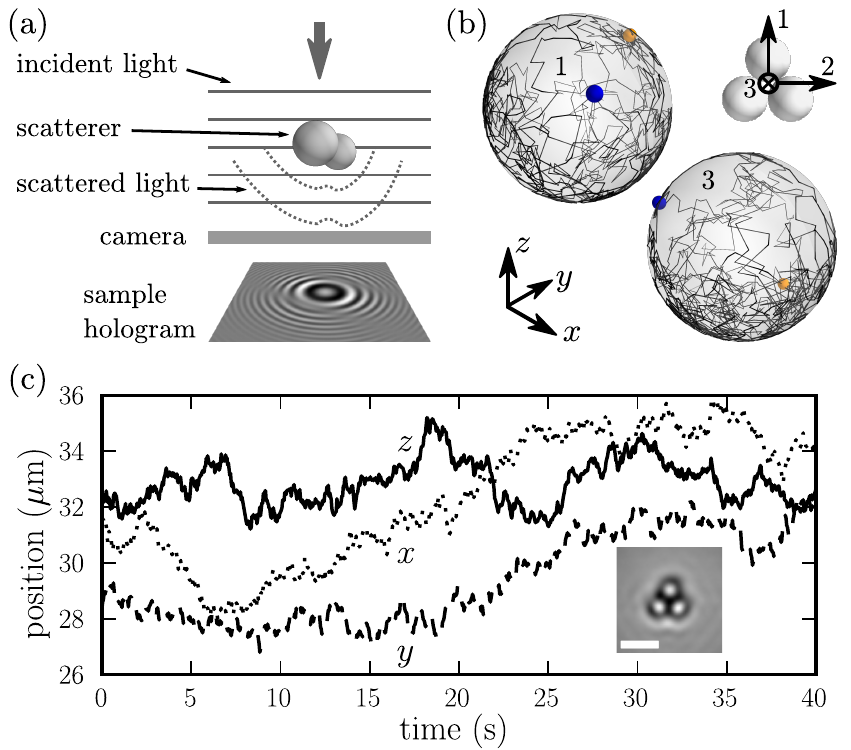}
\caption{\label{fig:schematic} (Color online) (a) Schematic overview
  of in-line holographic microscope used in the experiments.  (b) Axis
  trajectories for a diffusing triangular cluster of 1 $\mu$m diameter
  polystyrene spheres (inset: orientation of cluster axes). Renderings
  show the trajectories of axes 1 and 3 on a unit sphere. Blue (dark)
  markers indicate the start of the 40 s trajectory and orange (light)
  markers the end. (c) Laboratory-frame center-of-mass trajectories of
  the same triangular cluster over the same time interval as in
  (b). Inset: amplitude reconstruction~\cite{kreis_frequency_2002} of
  a recorded trimer hologram, showing the cluster structure. Scale
  bar, 2 $\mu$m.}
\end{figure}

In this Communication, we study the 3D diffusion of individual
colloidal clusters. We measure the diffusion tensor using a fast 3D
imaging technique, in-line digital holographic microscopy, which
involves recording a 2D hologram generated by interference between
light scattered from colloidal particles and the undiffracted,
transmitted beam
(Fig.~\ref{fig:schematic}(a))~\cite{lee_characterizing_2007}.  
Unlike 3D confocal 
stacks, 2D holograms can be recorded at sub-millisecond frame rates.
By fitting models based on electromagnetic scattering solutions to the
holograms~\cite{lee_characterizing_2007, fung_measuring_2011,
  fung_imaging_2012}, we recover the 3D dynamics of dimers and
triangular trimers of colloidal spheres. We resolve all the
translational and rotational components of the diffusion tensor to 1\%
precision or better. We experimentally demonstrate the effects of
asymmetry on diffusion, and even show that a small amount of
polydispersity results in symmetry breaking in the rotational
diffusion tensor components. Our measurements agree well with
numerical calculations and, more generally, show how diffusion tensors
can be measured in experimental systems relevant to self-assembly,
where theoretical predictions are challenging.

The diffusion tensor \tens{D} quantifies the translational and
rotational Brownian diffusion of an arbitrary rigid colloidal
particle. In general, six generalized coordinates $q^i$ (three
positions and three orientation angles) are needed to describe the
position and orientation of a rigid body. Although \tens{D} is
rigorously defined by generalizing Fick's law to an abstract ensemble
of particles diffusing in this 6-dimensional configuration space,
\tens{D} also describes correlations between displacements of the
$q^i$ for short lag times $\tau$~\cite{brenner_coupling_1967,
  harvey_coordinate_1980}\footnote{Eq.~\ref{eq:gen_correlation}
  strictly applies only for short $\tau$ because it assumes small
  changes in the generalized particle probability density arising from
  diffusion for short $\tau$~\cite{brenner_coupling_1967}}:
\begin{equation} \label{eq:gen_correlation}
\langle\Delta q^i \Delta q^j \rangle = 2D^{ij}\tau.
\end{equation} 
\tens{D}
may be partitioned into the following $3\times 3$ blocks:
\begin{equation}\label{eq:general_d}
\tens{D} = 
\begin{pmatrix}
\tens{D}^{tt} & \tens{D}^{\dagger tr} \\
\tens{D}^{tr} & \tens{D}^{rr} 
\end{pmatrix}.
\end{equation}
$\tens{D}^{tt}$ describes translational diffusion, $\tens{D}^{rr}$
rotational diffusion, and $\tens{D}^{tr}$ translation-rotation
coupling. The generalized Stokes-Einstein equation relates \tens{D} to
the friction tensor $\tens{\mathfrak{F}}$ describing the hydrodynamic
Stokes drag forces and torques on a moving particle: $\tens{D} = k_BT
\tens{\mathfrak{F}}^{-1}$~\cite{brenner_coupling_1967,
  harvey_coordinate_1980}. Predicting \tens{D} thus requires a
solution for the Stokes flow around a particle. Dimers are one of the
few non-spherical shapes for which analytical solutions exist; none
exist for trimers.

We make dimer and triangular trimer clusters through limited
aggregation of sulfate polystyrene spheres
(Invitrogen)~\cite{yake_fabrication_2006}, 1.3-$\mu$m diameter for
dimers and 1-$\mu$m diameter for trimers. We transfer these particles
into a 250 mM NaCl solution to screen the charge of the stabilizing
sulfate groups and start the aggregation, then we decrease the ionic
strength by quenching with deionized water (Millipore) after 1 minute
to arrest the aggregation. We then suspend the resulting mixture of
single particles, dimers, and larger clusters in a density-matched
solvent consisting of 50\% v/v D$_2$O and 50\% v/v H$_2$O with a salt
concentration of 1 mM.  We load the particles into sample cells made
from glass slides, coverslips, and 76-$\mu$m thick Mylar
spacers~\cite{fung_measuring_2011, fung_imaging_2012}. After finding a
cluster with the desired morphology using bright field microscopy, we
ensure that it is at least 30 $\mu$m away from sample cell walls or
other particles so that the cluster diffusion is unhindered by
boundaries~\cite{faucheux_confined_1994}. We record
holograms using an instrument previously described in the
literature~\cite{fung_imaging_2012, perry_real-space_2013} at a frame
rate of 25 frames per second and at ambient temperature $T = 296^{+2}_{-4}$ K. 

We then analyze the measured holograms to obtain 3D trajectories of
the clusters. We obtain particle sizes, refractive indices,
center-of-mass 3D positions, and three orientation angles (two for
dimers) by fitting an exact scattering solution to Maxwell's equations
to each recorded hologram~\cite{mackowski_calculation_1996,
  fung_measuring_2011}\footnote{See Supplementary Information for
  further details.}.
Figs.~\ref{fig:schematic}(b) and (c) show some
of the 3D data we obtain for a trimer.  We measure the components of
the translational block $\tens{D}^{tt}$ by directly applying
Eq.~\ref{eq:gen_correlation}, where the relevant $\Delta q^i$ are
relative to a coordinate system rigidly fixed to the
cluster~\cite{fung_measuring_2011, brenner_coupling_1967,
  harvey_coordinate_1980}. The correlation functions needed to measure
the diagonal components of $\tens{D}^{tt}$ are cluster-frame
mean-squared displacements (MSDs). To measure the components of
$\tens{D}^{rr}$, we examine the dynamics of the axis vectors
$\vect{u}_i$ fixed to a cluster.  The tips of the $\vect{u}_i$ diffuse
along the surface of a unit sphere, as illustrated in
Fig.~\ref{fig:schematic}(b). We compute autocorrelations of the
$\vect{u}_i$, which are related to $D_{r,i}$, the 3 diagonal components of
$\tens{D}^{rr}$, as follows~\cite{favro_theory_1960}:
\begin{equation}\label{eq:axis_autocorr}
\langle \vect{u}_i(t) \cdot \vect{u}_i(t+\tau) \rangle = 
\exp \left[ \left( D_{r,i} - \sum_j D_{r,j} \right) \tau \right].
\end{equation}
For the axisymmetric dimer, we consider a related quantity, the MSD of
the axis unit vector \vect{u}~\cite{doi_theory_1988,
  mukhija_translational_2007, cheong_rotational_2010,
  fung_measuring_2011}:
\begin{equation}\label{eq:dimer_axis}
\langle \Delta \vect{u}^2 (\tau) \rangle = 2 \left( 1 - \langle
  \vect{u}(t) \cdot \vect{u}(t + \tau) \rangle \right).
\end{equation}
% I think this makes the relationship clearer, and also keeps us from
% having to explicitly define D_r here.

\begin{figure}
\includegraphics{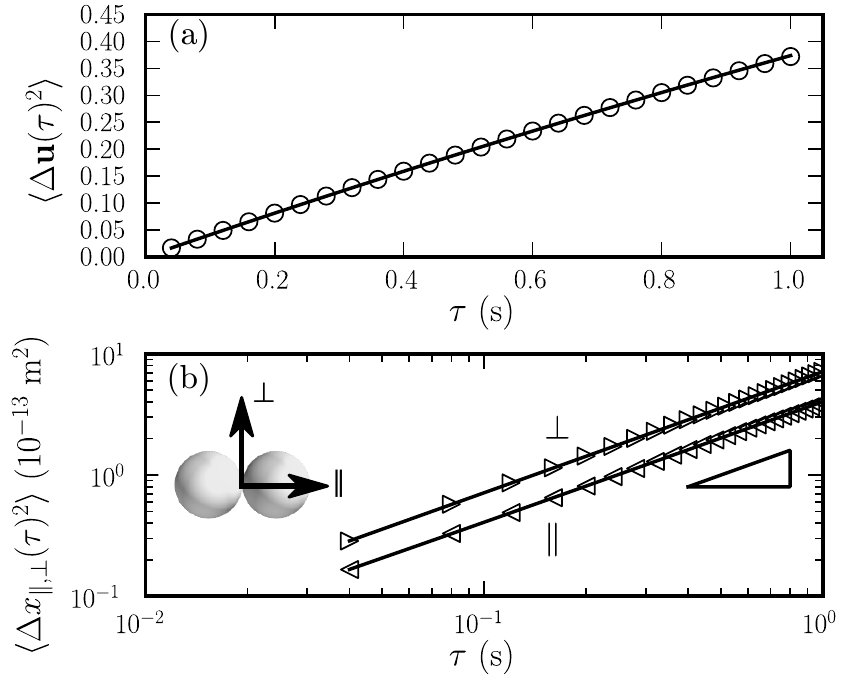}
\caption{\label{fig:dimer} (a) Axis MSD for dimer of 1.3-$\mu$m
  diameter spheres. Open symbols are measurements; solid line is the
  best fit to Eq.~\ref{eq:dimer_axis}, where $\langle \vect{u}(t)
  \cdot \vect{u}(t+\tau) \rangle = \exp(-2D_{r,\perp}\tau)$.
  (b) Cluster-frame MSD computed
  for the same dimer as in (a). Open symbols are measurements; solid
  lines are linear fits. Triangle shows slope of 1. Error bars are
  calculated using a block decorrelation
  technique~\cite{flyvbjerg_error_1989}; they are comparable in size
  to the plotting symbols or smaller. Inset: orientation of parallel
  $(\parallel)$ and perpendicular $(\perp)$ axes. }
\end{figure}

We find good agreement between \tens{D} for a dimer, measured to 0.5\%
precision \footnote[2]{} from a time series of 22,000 holograms, 
and analytical and numerical predictions. The measured axis MSD, along
with a best fit to Eq.~\ref{eq:dimer_axis}, and the cluster-frame MSDs
are shown in Fig.~\ref{fig:dimer}.  Dimers of two identical spheres
have three mirror planes and an axis of continuous rotational
symmetry, so in the coordinate system shown in the inset of
Fig.~\ref{fig:dimer}(b), \tens{D} is diagonal. But due to the breaking
of spherical symmetry, \tens{D} has four rather than two unique
elements: the translational diffusion constants $D_\parallel$ and
$D_\perp$, and the rotational diffusion constants $D_{r,\parallel}$ and
$D_{r,\perp}$~\cite{fung_measuring_2011, brenner_coupling_1967}. The
subscripts denote translations along, or rotations about, the dimer
symmetry axis ($\parallel$) and the two degenerate perpendicular axes
($\perp$). Because we cannot observe rotations about the dimer axis,
we can only measure $D_{r,\perp}$. Moreover, we can only measure a
combined MSD, $\langle \Delta x_\perp(\tau)^2\rangle$, along the
perpendicular axes. We then extract $D_\parallel$ and $D_\perp$ from
linear fits to the MSDs. $\langle \Delta x_\parallel^2 (\tau)\rangle$
has a slope of $2D_\parallel$ in accordance with
Eq.~\ref{eq:gen_correlation}, and $\langle \Delta x_\perp^2
(\tau)\rangle$ has a slope of $4D_\perp$. Our measurement of
$D_\parallel/D_\perp$, which is a universal constant for any Brownian
dimer, agrees well with predictions from the shell modeling code
\textsc{hydrosub}~\cite{garcia_de_la_torre_hydrodynamic_2002} and the
exact Stokes solution of Nir \& Acrivos~\cite{nir_creeping_1973}
(Table~\ref{tab:dimer_results}).

The individual elements of \tens{D} also agree with calculations, once
we account for the sphere radius $a$ and the solvent viscosity $\eta$.
We determine an effective sphere radius $a_{\mathrm{eff}}$ from the
measured ratios $D_\parallel/D_{r,\perp}$ and $D_\perp/D_{r,\perp}$,
which depend only on $a$~\cite{nir_creeping_1973}. Then we determine
the best-fit $\eta_{\mathrm{eff}}$ from the measured elements of
\tens{D}. We find an effective $a_\mathrm{eff} = 709$ nm, which is
larger than the optical radius $a_\mathrm{opt} = 650$ nm obtained by
fitting the holographic data. The larger effective radius is
consistent with typical dynamic light scattering measurements of the
size of colloidal spheres, which show enhanced hydrodynamic radii due
to charge or hairy surface layers on the
particles~\cite{seebergh_evidence_1995,
  gittings_determination_1998}. The best-fit viscosity is
$\eta_\mathrm{eff}=1.159$ mPa\,s, consistent with measurements of the
diffusion constant of a freely diffusing sphere in the same sample,
which acts as an \textit{in situ} thermometer, combined with bulk
viscosity measurements using a Cannon-Manning capillary viscometer;
these together give a solvent viscosity of $1.19\pm0.04$ mPa\,s.  We
use this procedure because the solvent viscosity has a strong
temperature dependence \footnote[2]{}.  The elements of \tens{D}
computed with these effective parameters agree with our measurements
to better than 1\% (Table~\ref{tab:dimer_results}).  We also note that
the \textsc{hydrosub} prediction for $D_{r,\perp}$ differs from the
analytical prediction by about 1\%, which is consistent with prior
studies~\cite{garcia_de_la_torre_improved_2007}. The agreement between
our measurements and the analytical prediction suggests that our
measurement accuracy is at least comparable to, if not better than,
that of \textsc{hydrosub}.

\begin{table}
  \caption{\label{tab:dimer_results} Measured diffusion tensor
    elements for dimer in Fig.~\ref{fig:dimer}, along with analytical
    calculations from an exact Stokes solution
    \cite{nir_creeping_1973} and numerical calculations from
    \textsc{hydrosub}~\cite{garcia_de_la_torre_hydrodynamic_2002}. Calculations
    use a best-fit particle radius $a_\mathrm{eff} = 709$ nm and
    solvent viscosity $\eta_\mathrm{eff} = 1.159$ mPa\,s. Experimental
    uncertainties determined from best fits in Fig.~\ref{fig:dimer};
    see Supplementary Information for details.}   
\begin{ruledtabular}
\begin{tabular}{l  c  c  c }
  & Experiment & Exact & \textsc{hydrosub} \\ \hline
\rule{0pt}{3ex}$D_{r,\perp}$ (s$^{-1}$) & 0.1034 $\pm$ 0.0006 & 0.1034  &
0.104 \\
$D_\parallel$ ($\times 10^{-13}$ m$^2$s$^{-1}$) & 2.015 $\pm$ 0.012 &
2.010 & 2.02 \\
$D_\perp$ ($\times 10^{-13}$ m$^2$s$^{-1}$) & 1.785 $\pm$ 0.007 &
1.790 & 1.80 \\
$D_\parallel/ D_\perp$ & 1.129 $\pm$ 0.011 & 1.123 & 1.12
\end{tabular}
\end{ruledtabular}
\end{table}

\begin{figure}
\includegraphics{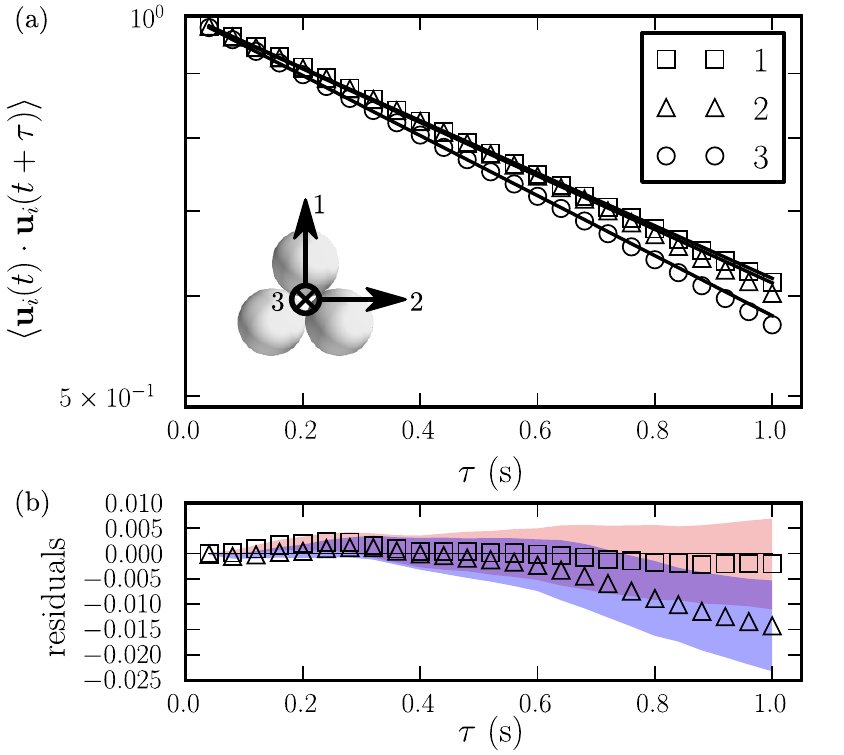}
\caption{\label{fig:trimer_rot} (Color online) (a) Cluster axis
  autocorrelations $\langle \vect{u}_i(t) \cdot \vect{u}_i(t + \tau)
  \rangle$ for a trimer of 1-$\mu$m diameter spheres, showing
  anisotropic rotational diffusion. Open symbols are experimental
  measurements; error bars are comparable to or smaller than
  symbols. Solid lines are fits to exponential decays. Inset shows
  cluster axis orientation.  (b) Residuals for fits of a single
  exponential decay to the in-plane axis autocorrelations ($i = 1$ and
  $2$). Solid line indicates best fit exponential. Red (light) and
  blue (dark) shaded regions denote error bars.}
\end{figure}

\begin{figure}
\includegraphics{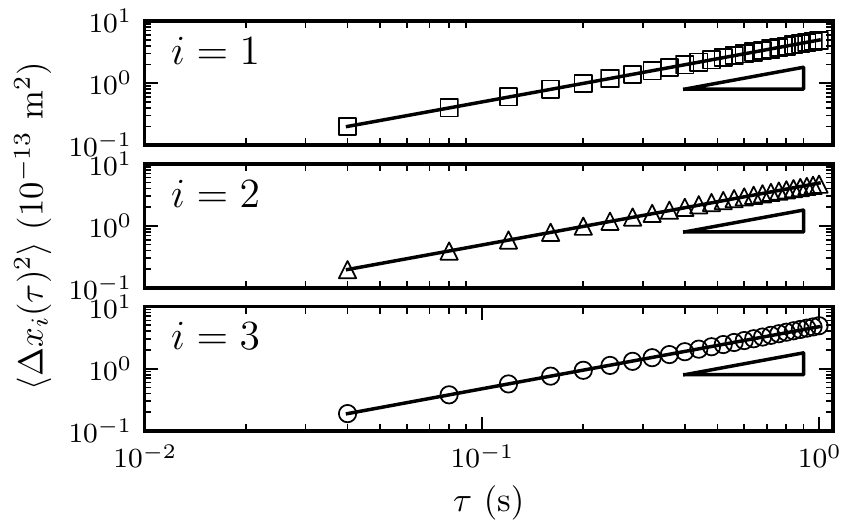}
\caption{\label{fig:trimer_trans} Body-frame MSDs for the same trimer
  in Fig.~\ref{fig:trimer_rot}. See inset in
  Fig.~\ref{fig:trimer_rot}(a) 
for axis orientations $i$. Open symbols are experimental measurements;
error bars are comparable to or smaller than symbols. Solid lines are
linear fits. Triangles show MSD slope of 1.}
\end{figure}

Measurements on trimers reveal anisotropic translational and
rotational diffusion.  Trimers of identical particles have two mirror
planes, making $\tens{D}^{tt}$ and $\tens{D}^{rr}$
diagonal~\cite{happel_low_1965, brenner_coupling_1967} in the
coordinate system shown in the inset of Fig.~\ref{fig:schematic}(b). 
We denote the six diagonal elements as $D_{t,1}$, $D_{t,2}$, $D_{t,3}$,
$D_{r,1}$, $D_{r,2}$, and $D_{r,3}$. In contrast to dimers, trimers
lack axisymmetry, allowing us to observe rotations about all three
axes and measure all six elements.  In Fig.~\ref{fig:trimer_rot}(a), we
show the axis autocorrelations $\langle
\vect{u}_i\cdot\vect{u}_i(t+\tau) \rangle$ computed from 20,000
holograms, as well as best fits to exponential decays. The
autocorrelation of axis 3 decays more rapidly than the
autocorrelations of axes 1 and 2, in agreement with expectations: as
shown in Eq.~\ref{eq:axis_autocorr}, $\langle \vect{u}_3(t) \cdot
\vect{u}_3(t + \tau) \rangle$ depends on $D_{r,1}$ and $D_{r,2}$, both
of which should be larger than $D_{r,3}$ due to hydrodynamics.  The
elements of the diffusion tensor that we extract from this data are
shown in Table~\ref{tab:trimer_results}. The difference between
$D_{r,3}$ and both $D_{r,1}$ and $D_{r,2}$ is much larger than the
experimental uncertainty, showing clear evidence for anisotropic
rotational diffusion. The translational diffusion we observe is
similarly anisotropic (Fig.~\ref{fig:trimer_trans} and
Table~\ref{tab:trimer_results}).

Interestingly, although our measurements of the dimensionless ratios
$D_{r,1}/D_{r,3}$ and $D_{t,1}/D_{t,3}$ agree well with the
\textsc{hydrosub} predictions, we observe small but statistically
significant differences between the elements of \tens{D} corresponding
to the two in-plane axes 1 and 2. Figure \ref{fig:trimer_rot}(b)
indicates that the autocorrelations of axes 1 and 2 are poorly fit by
a single exponential decay. If the particles in the trimer are
identical, the threefold symmetry axis of the trimer ensures that
$D_{t,1} = D_{t,2}$ and $D_{r,1} = D_{r,2}$
\footnote{See~\cite{happel_low_1965, brenner_coupling_1967}. Brenner
  does not explicitly treat discrete rotational symmetry, but his
  symmetry arguments are readily applied to this case.}.  Thus the
differences between these elements of the tensor imply that the
particles in our trimer are not in fact identical. We performed
\textsc{hydrosub} calculations to confirm that weakly breaking
threefold symmetry results in differences between the in-plane
elements of \tens{D}. Our measured ratio $D_{r,1}/D_{r,2} = 1.03 \pm
0.02$ corresponds to a 3\% size difference between the spheres. This
is consistent both with the particle manufacturer's certificate of
analysis as well as with particle size differences determined from
holograms.  This shows that even a small amount of particle
polydispersity can break the threefold rotational symmetry to a
measurable degree.

\begin{table}
  \caption{\label{tab:trimer_results} Measured diffusion tensor elements
    for trimer shown in Figs.~\ref{fig:trimer_rot} and
    \ref{fig:trimer_trans} with comparisons to 
    computations from
    \textsc{hydrosub}~\cite{garcia_de_la_torre_hydrodynamic_2002}. Computations 
    use $a = 500$ nm obtained optically from the best-fit hologram
    models and $\eta = 1.049$ mPa\,s from 
    single-particle diffusion data; the difference in $\eta$ from the
    dimer measurements is due to a difference in room
    temperature. Experimental uncertainties determined from best fits
    in Figs.~\ref{fig:trimer_rot} and \ref{fig:trimer_trans}; see
    Supplementary Information for details.} 
\begin{ruledtabular}
\begin{tabular}{l c c}
 & Experiment & \textsc{hydrosub} \\ \hline
\rule{0pt}{3ex}$D_{r,1}$ (s$^{-1}$) & 0.278 $\pm$ 0.002 & 0.296 \\
$D_{r,2}$ (s$^{-1}$) & 0.270 $\pm$ 0.002 & 0.296 \\
$D_{r,3}$ (s$^{-1}$) & 0.210 $\pm$ 0.002 & 0.220 \\
$D_{r,1} / D_{r,3}$ & 1.32 $\pm$ 0.02 & 1.34 \\
$D_{r,1} / D_{r,2}$ & 1.03 $\pm$ 0.02 & 1.00 \\
$D_{t,1}$ ($\times 10^{-13}$ m$^2$s$^{-1}$) & 2.466 $\pm$ 0.015 & 2.64
\\
$D_{t,2}$ ($\times 10^{-13}$ m$^2$s$^{-1}$) & 2.446 $\pm$ 0.015 & 2.64
\\
$D_{t,3}$ ($\times 10^{-13}$ m$^2$s$^{-1}$) & 2.372 $\pm$ 0.015 & 2.41 \\
$D_{t,1}/D_{t,3}$ & 1.04 $\pm$ 0.01 & 1.09 
\end{tabular}
\end{ruledtabular}
\end{table}

Overall, our work demonstrates experimentally how both large and small
differences in the symmetry of small particles affect the diffusion
tensor. The technique we use, holographic microscopy, can measure
elements of the diffusion tensor to high precision, $1$\% or better,
small enough to resolve weak symmetry breaking due to particle
polydispersity. The high precision is enabled by the inherently short
acquisition times of the technique, allowing us to study rapidly
diffusing systems which we can image for up to hundreds of rotational
diffusion times $1/D_{r,i}$. These longer timescales more clearly
reveal anisotropy in the diffusion tensor. Although here we have
measured the diffusion of isolated clusters, it should be possible to
use the same technique to measure diffusion tensors in environments
that are relevant to self-assembly but challenging for computation. In
particular, it may be possible to measure diffusion tensors near
boundaries or other particles, or for clusters that have internal
degrees of freedom. The few studies examining the effect of
interparticle hydrodynamic couplings on particle diffusion have been
restricted to spheres in planar
geometries~\cite{crocker_measurement_1997, merrill_many-body_2010,
  lele_colloidal_2011}. Furthermore, measurements on other, even less
symmetric clusters may be able to reveal translational-rotational
coupling as well as the off-diagonal rotational and translational
elements in the diffusion tensor.

%\appendix
%\section{}

\begin{acknowledgments}
  We thank Jos\'e Garc\'ia de la Torre, Daniela J.~Kraft, Thomas
  G.~Dimiduk, and Rebecca W.~Perry for helpful discussions. This work
  was supported by the National Science Foundation through CAREER
  grant no.~CBET-0747625 and through the Harvard MRSEC, grant
  no.~DMR-0820484. We also acknowledge support from the Kavli
  Institute for Bionano Science \& Technology at Harvard.
\end{acknowledgments}

\bibliography{fung_diffusion}

%\appendix

\section*{Supplementary Information}

% label figs and equations with S1, S2, etc
\renewcommand{\thefigure}{S\arabic{figure}}
\renewcommand{\theequation}{S\arabic{equation}}
\setcounter{figure}{0} % reset counter
\setcounter{equation}{0}
\setcounter{section}{0}

\section{Hologram Analysis and Data Reduction}
We describe in further detail how we fit scattering models to the
holograms we record of dimers and trimers to extract 3D dynamical
information, with attention to issues that arise with the large number
of holograms ($\sim$20,000) we must analyze for each case.

\subsection{Model Fitting Procedure}
Our technique of fitting scattering solutions to holograms has
previously been described \cite{fung_measuring_2011,
  fung_imaging_2012}. For the dimers, we fit a scattering model that
depends on one refractive index for both particles, the radius of each
particle, the 3D position of the dimer center of mass, 2 orientational
Euler angles, and a scaling parameter \cite{fung_measuring_2011}. The
model for trimers differs only in that we fit for three orientation
angles and for only one average radius \cite{fung_imaging_2012}. 

The largest bottleneck in fitting scattering models to large numbers
of holograms using Levenberg-Marquardt minimizers is that an initial
guess for the model parameters needs to be provided for each
hologram. With time series of holograms, only one initial guess is
necessary, in principle: we can use the best-fit parameters of each
hologram as the initial guess for the next. We have found it effective
to speed up this process by doing an initial rough fit to a randomly
chosen subset of 10\% of the pixels of each hologram. Subsequently, we
use the parameters obtained from the rough fits as initial guesses
for a fit to all the pixels. We do this second stage of fitting in
parallel.

\subsection{Validating Model Fits to Trimer Holograms}
\label{sec:validation}
The model for the trimer holograms has an additional orientational
degree of freedom compared to the model for the dimer holograms. We
have noticed that on occasion the fitter converges to best-fit
parameters that result in the best-fit model hologram having subtle
differences when compared to the experimental hologram; this usually
stems from the orientation angles being incorrect. We do not observe
this problem for the dimer holograms. To detect holograms with
potentially incorrect best-fit parameters, we inspect the $R^2$
statistic \cite{fung_imaging_2012} of the fits. We also compute a
$\chi^2$ statistic for a binary version of the experimental and
best-fit holograms, where all pixels above the mean of 1 are set to a
value of 1 and all remaining pixels are set to a value of 0. The
binary image is much more sensitive to the shape of the interference
fringes. 

When we compute correlation functions such as mean-squared
displacements from the trimer holograms, we reject the contribution
from any holograms where either $R^2$ or binary $\chi^2$ is worse than
2 standard deviations from a rolling mean. Manual inspection of 200
randomly chosen trimer holograms that were not rejected under these
criteria revealed 7 questionable fits. We infer from the Poisson
distribution that, to a 99\% confidence level, the percentage of
remaining bad fits is less than 8\%. We also reject the contribution
from a given pair of holograms if the probability of obtaining either
a center of mass displacement or angular displacement of the observed
magnitude is less than $10^{-5}$. We compute these probabilities using
estimates for the diffusion tensor elements, and choose the threshold
of $10^{-5}$ to avoid biasing the observed distribution and to make
the cutoffs weakly sensitive to the estimates for \tens{D}.

Performing this cutoff procedure requires knowing the probability
distributions governing translational and rotational displacements.
The probability distribution for translational displacements is
Gaussian, but the distribution function for rotational displacements is
not. Instead, the probability density function
$f_i(\theta;\,\tau)$ for observing an angle $\theta$ between cluster
axis $\vect{u}_i(t)$ at a given time $t$ and after a time interval
$\tau$ is given by
\begin{equation}\label{eq:rot_dist}
f_i(\theta; \tau) = \sum_{\ell = 0}^\infty Y_\ell^0(0)Y_\ell^0(\theta)
\exp\left[-\ell(\ell+1)D_{r,\mathrm{eff}}\tau \right]
\end{equation}
where $D_{r,\mathrm{eff}} = (D_{r,j} + D_{r,k})/2$, $D_{r,j}$ and
$D_{r,k}$ are the elements of $\tens{D}^{rr}$ describing rotations
about the two cluster axes other than $i$, and $Y_\ell^0(\theta)$
denote spherical harmonics with $m=0$. We briefly discuss the
origin of this distribution function in Section~\ref{sec:rot_dist}.

\begin{figure}
\centering
\includegraphics{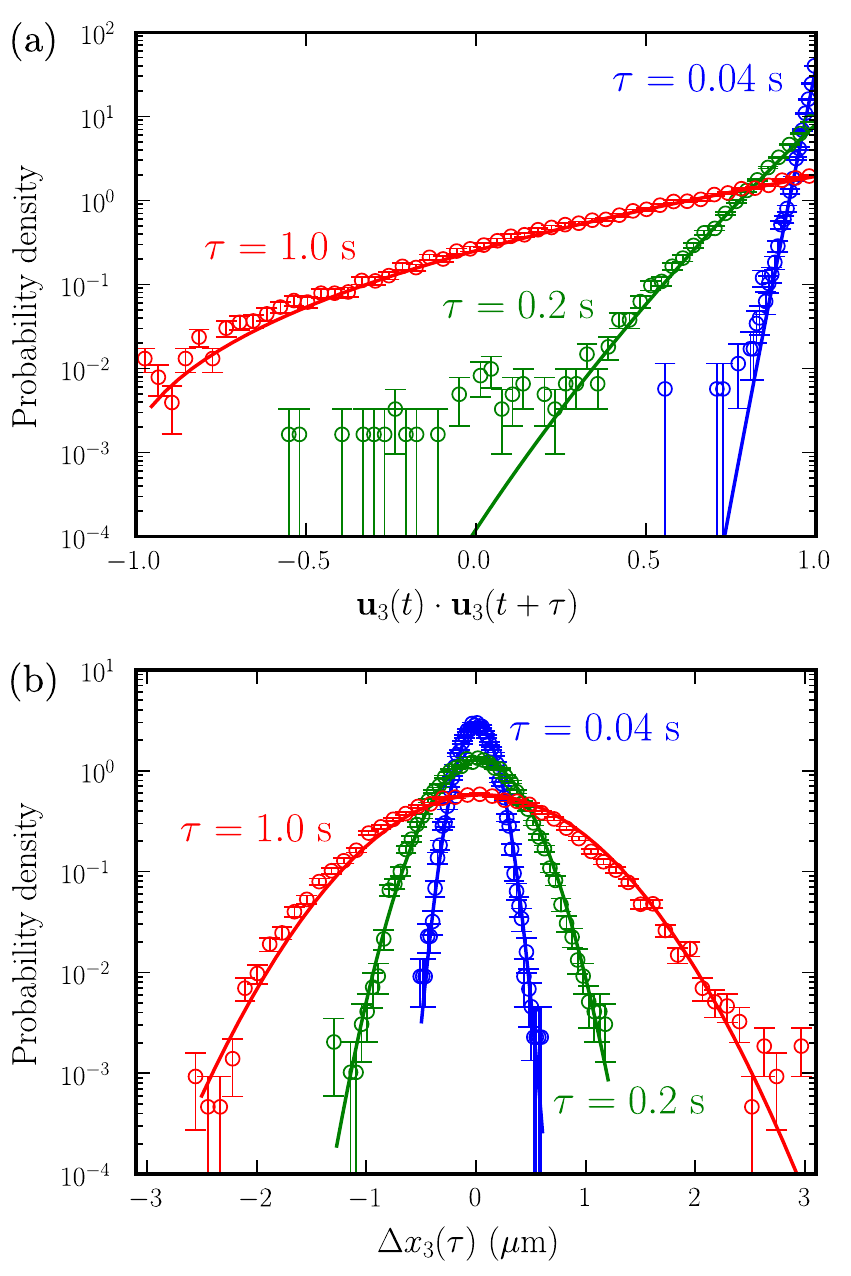}
\caption{\label{fig:vanhove} Distribution functions for trimer angular
displacements and cluster-frame displacements. Histogram points
computed from experimental data are shown in open symbols; solid lines
show theoretical predictions computed from elements of \tens{D}
reported in Table II of the body of the paper. (a) Rotational dynamics
of $\vect{u}_3$. Predicted distribution computed from
Eq.~\ref{eq:rot_dist}. (b) Cluster-frame displacements along axis
3. Theoretical distribution is a Gaussian with a mean of 0 and a
variance of $2D_{t,3}\tau$. }
\end{figure}

% discuss van hoves
As a final verification that our holographic imaging is correct and
that any remaining errors do not substantially affect the dynamics we
measure, we compute probability distribution functions for the
dynamical quantities we use to measure \tens{D} from the data. Figure
\ref{fig:vanhove} shows a representative sample for several lag times
$\tau$. We first examine the
cosine of the angle traversed by $\vect{u}_3$, or $\vect{u}_3(t)\cdot
\vect{u}_3(t + \tau)$, in Figure \ref{fig:vanhove}(a). Aside from a
noise floor, we find that the measured distributions agree well with
the expected distribution computed from Eq.~\ref{eq:rot_dist} and the
measured values of $\tens{D}^{rr}$. We observe similarly good
agreement for the distribution of particle-frame displacements along
axis 3 shown in Figure \ref{fig:vanhove}(b).

\subsection{Estimating Uncertainties in Elements of \tens{D}}

The uncertainties in the elements of \tens{D} are derived from the
cluster trajectories obtained from holographic microscopy. The
elements of \tens{D} are measured by computing a correlation function
(an MSD or axis autocorrelation) for a range of time steps $\tau$, and
by fitting a theoretical expression involving the elements of \tens{D}
(Eqs.~1, 3, or 4 in the manuscript) to the points of the correlation
function. The uncertainties we report are the appropriate diagonal
elements of the covariance matrix of the best-fit parameters.

Determining the uncertainties in the elements of \tens{D} in this
manner requires the points in the correlation functions to be weighted
by the uncertainty associated with each point. The uncertainty we
consider is the statistical error related to the number of
uncorrelated displacements that we use to calculate each point. The
statistical error is directly estimated from the appropriate cluster
trajectory using a block decorrelation procedure
\cite{flyvbjerg_error_1989}. We choose the block decorrelation method
because it does not require any \textit{a priori} assumptions about
the underlying statistical distribution governing the displacements.

\subsection{Comparison to Confocal Microscopy}

The precision with which we measure elements of \tens{D} consequently
increases with the number of observed displacements and hence with the
length of the trajectories we observe. The rapid acquisition times of
holographic microscopy give it an advantage over complementary 3D
imaging techniques such as confocal microscopy in that a considerably
larger number of 3D images can be acquired in the same amount of
experimental time.

The main advantage of holographic microscopy over confocal microscopy,
however, lies in the greater sensitivity of experiments using
holographic microscopy to weakly anisotropic diffusion. In confocal
experiments on diffusion, the acquisition time needed to scan through
a 3D volume ($\sim 1$ s or more) requires the dynamics to be slowed
down through the use of larger particles and more viscous solvents.
This results in the elements of \tens{D} being much smaller. For
example, tetrahedral sphere clusters used in confocal measurements of
diffusion~\cite{hunter_tracking_2011} have an isotropic rotational
diffusion constant of $D_r \sim 5 \times 10^{-3}$ s$^{-1}$, nearly two
orders of magnitude smaller than in our trimer
experiment. Consequently, given the same amount of experimental time,
confocal experiments access much shorter timescales relative to the
rotational diffusion times than holographic experiments. This makes it
more challenging to observe statistically significant anisotropy in
\tens{D}, as we now show.

Demonstrating anisotropic diffusion requires showing that the ratio of
the rotational autocorrelation functions about axes $i$ and $j$
differs from 1 by a statistically significant amount. Eq.~3 of our
manuscript gives this ratio in terms of the relevant elements of
\tens{D}:
\begin{align}
\frac{\langle \vect{u}_i(t) \cdot \vect{u}_i(t+\tau) \rangle}{\langle
  \vect{u}_j(t) \cdot \vect{u}_j(t+\tau) \rangle} &=
\exp\left[-(D_{r,j} - D_{r,i})\tau \right] \\
&\approx 1 - (D_{r,j} - D_{r,i})\tau + \ldots
\end{align}
where we have assumed in the second step that the anisotropy is
small. The ratio differs from 1 in proportion to the \emph{magnitude}
of the difference in the rotational diffusion constants, rather than
in proportion to the relative difference.  Consider a confocal
experiment and a holographic experiment on systems with the same
relative anisotropy $D_{r,i}/D_{r,j}$, where both experiments measure
the same number of independent displacements over the same time
interval $\tau$. Both experiments will compute autocorrelations at
$\tau$ with the same precision and will require similar amounts of
experimental time. But for the confocal experiment, $(D_{r,j} -
D_{r,i})\tau$ will be smaller, and may even be comparable to the
measurement precision of the autocorrelations. Thus, because
holographic microscopy can study more rapidly diffusing clusters, it
is easier to observe weakly anisotropic diffusion, as we show in our
measurement of the the 3\% difference between $D_{r,1}$ and $D_{r,2}$
for the trimer.

\section{Probability Distribution for Rotational Displacements}
\label{sec:rot_dist}
We briefly describe the origin of Eq.~\ref{eq:rot_dist}, the
probability density function for finite rotational displacements. As
described in Section \ref{sec:validation}, we use this distribution to
reject pairs of holograms that exhibit highly improbable angular
displacements, most likely due to an incorrect model fit. Our
discussion here is primarily physical and draws heavily from that of
Berne \& Pecora \cite{berne_dynamic_1976}; the reader interested in a
more rigorous but abstract discussion is referred to
\cite{favro_theory_1960, brenner_coupling_1967}.

To calculate the relevant distribution function, we will consider what
happens to an imaginary ensemble of clusters undergoing rotational 
diffusion. 
We will assume that translation-rotation coupling is negligible, so
that we can consider the rotational motions independently of the
translations. Suppose that we observe only the motion
of one body axis $\vect{u}_i$ as the clusters in the ensemble undergo
rotational diffusion. Lastly, suppose that that we prepare the
ensemble
such that at $t=0$, $\vect{u}_i$ for every cluster lies at the same
point on the unit sphere, which we may choose to be at $\theta=0$
without loss of generality. We seek to compute the probability
distribution $f_i(\theta, \phi;\,\tau)$ such that
\begin{equation}
\int_{\phi_0}^{\phi_1} \int_{\theta_0}^{\theta_1}
f_i(\theta,\phi;\,\tau) \,\sin\theta d\theta d\phi
\end{equation}
gives the probability of finding $\vect{u}_i$ between $\phi_0$ and
$\phi_1$ and between $\theta_0$ and $\theta_1$ at $t=\tau$.

For isotropic rotational diffusion, such as that of a
sphere, calculating $f_i$ is straightforward: the diffusion can be
described by a rotational Fick's law 
characterized by a single rotational diffusion constant $D_r$
\cite{berne_dynamic_1976}. The initial condition
\begin{equation}
f_i(\theta,\phi;0) = \frac{\delta(\theta)}{2\pi\sin\theta},
\end{equation}
where $\delta(\theta)$ denotes the Dirac delta function, then
determines $f_i(\theta,\phi;\,\tau)$. This idea can be generalized 
to the case we are interested in, where
$\tens{D}^{rr}$ is diagonal but not isotropic, using Brenner's
tensorial formalism \cite{brenner_coupling_1967}. Because the details
are quite involved\footnote{Displacements of the 3 orientational
  generalized coordinates $q^i$ in Brenner's formalism cannot describe
  non-infinitesimal rotations. To describe finite rotations, it is
  necessary to adopt a formalism such as Euler angles; see Sec.~IX of
  Brenner's paper for details. The general case may also be handled by
  coordinate-free operator methods \cite{favro_theory_1960}.},
we instead give a physical argument that allows us
to apply the isotropic solution to the anisotropic case. 

The time evolution of $f_i$ must be governed by
$D_{r,j}$ and $D_{r,k}$, the diffusion constants for rotations about
the other two cluster axes. In our ensemble of clusters, prepared such 
that all clusters initially have $\vect{u}_i$ at $\theta = 0$, the
clusters will not all have the same
orientation: $\vect{u}_j$ and $\vect{u}_k$ can lie anywhere on the
equator of the unit sphere. Consequently, observing only the motion of
$\vect{u}_i$, we will on average observe $f_i$ evolving according
to an effective rotational diffusion constant $D_{r,\mathrm{eff}}$
where $D_{r,\mathrm{eff}} = (D_{r,j} + D_{r,k})/2$. 
We may then straightforwardly adopt the result
from isotropic diffusion \cite{berne_dynamic_1976}, which leads to
Eq.~\ref{eq:rot_dist}.
Note that $f_i$ is ultimately independent of
$\phi$ because of the symmetric manner in which we prepared the
ensemble. 

\section{Estimating Solvent Viscosities}

Here we discuss in greater detail the inference of the solvent
viscosities $\eta$ needed to compare experimentally measured elements
of the diffusion tensor \tens{D} to theoretical predictions.

\begin{figure}
\centering
\includegraphics{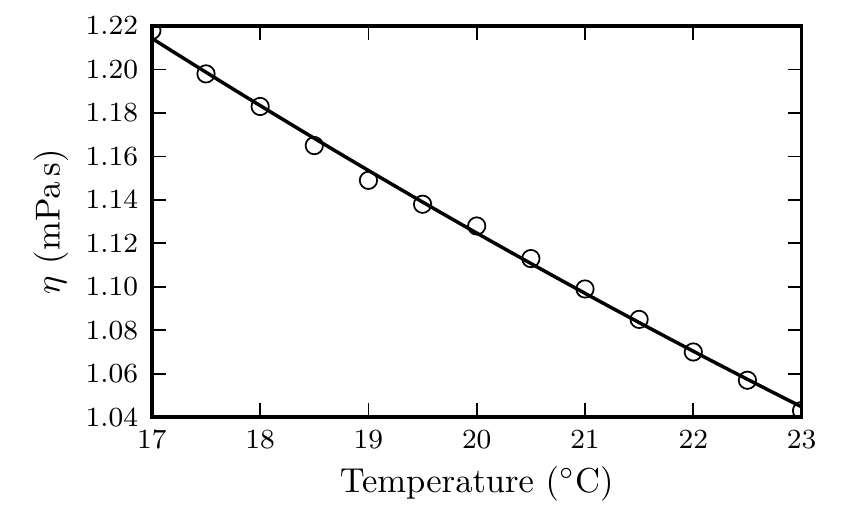}
\caption{\label{fig:visc} Temperature dependence of solvent
  viscosity. Data points, open symbols, were measured with a
  Cannon-Manning capillary viscometer. The solid line is a best-fit
  quadratic function that allows for interpolation between the measured
  points. Ambient laboratory temperatures in the diffusion experiments
  varied from 19--23 $^\circ$C.}
\end{figure}

The viscosity of our solvent, a mixture of H$_2$O and D$_2$O chosen to
density-match polystyrene particles, has a strong temperature
dependence. Figure~\ref{fig:visc} shows that the viscosity of a bulk
sample of the solvent, measured using a Cannon-Manning capillary
viscometer, varies by nearly 20\% over a 6$^\circ$C temperature
range. We performed all the experiments described in the body of the
paper at room temperature, but we have observed that the room
temperature can change by several $^\circ$C over the course of a few
hours, most likely due to the cycling of the building heating and air
conditioning systems. Moreover, particularly if the laboratory room
temperature is changing, the temperature in the sample, sealed in a
glass sample cell, may differ from that of the surrounding air.

Consequently, we believe that that the best way to estimate the solvent
viscosity is to observe the \textit{in situ} diffusion of single
colloidal spheres, which are always present in the sample due to the
arrested aggregation technique we use to make the clusters, either
immediately before or immediately after imaging the diffusion of a
cluster of interest. The Stokes-Einstein relation gives the
translational diffusion constant\footnote{In terms of the diffusion
  tensor \tens{D}, for a sphere \tens{D} is
  diagonal, and $\tens{D}^{tt} = D\tens{I}$, where \tens{I} is the
  identity tensor.} $D$ in terms of the temperature $T$,
the particle radius $a$, and the solvent viscosity $\eta$:
\begin{equation}\label{eq:stokes-einstein}
D = \frac{k_BT}{6\pi\eta a}.
\end{equation}
Using Eq.~\ref{eq:stokes-einstein}, once we determine $D$ for a
diffusing sphere of radius $a$, we can infer the ratio
$k_BT/\eta$. Because of the strong temperature dependence illustrated
in Figure \ref{fig:visc}, $k_BT$ and $\eta$ should not be viewed as
independent parameters. Moreover, from dimensional considerations,
the elements of \tens{D} are always proportional to $k_BT/\eta$. Once
we determine $k_BT/\eta$, we use the best-fit line to the data in
Figure \ref{fig:visc} to infer $\eta$ and $k_BT$ separately. While
this is not the usual context in which microrheological experiments 
are performed, we essentially treat the diffusing single spheres
as \textit{in situ} thermometers.

We obtain $D$ from an MSD computed from the 3D trajectory of a
diffusing particle: $\langle \Delta \vect{r}^2(\tau) \rangle =
6D\tau$. In all cases, we obtain the trajectory using holography and
record holograms at 25 frames per second. We obtain a radius, index of
refraction, and 3D position from each hologram by fitting a model
based on the Lorenz-Mie solution \cite{lee_characterizing_2007}.

For the dimer experiment, which used particles with a nominal radius
of 650 nm, we measure $D=2.533\pm0.017 \times 10^{-13}$
m$^2$s$^{-1}$ for a diffusing particle with an optical radius of 639
nm. If we assume that the particle has the same enhanced hydrodynamic
radius of 709 nm as we inferred from the dimer data, independent of
any considerations of $k_BT$ or $\eta$, we can subsequently use the
data in Figure \ref{fig:visc} to infer a solvent viscosity of 1.187
mPa\,s, which is within 3\% of the best-fit solvent viscosity, 1.159
mPa\,s. The consistency of these values, along with the excellent
agreement between the measured and predicted values of
$D_\parallel/D_\perp$, which is independent of $a$ and $k_BT/\eta$,
validates our dimer measurements.

For the trimer experiment, we measured $D=3.996\pm0.055 \times 10^{-13}$
m$^2$s$^{-1}$ for a diffusing sphere of nominal radius 500 nm. 
With no analytical theory as we had for dimers, we cannot
rigorously find a best-fit radius for the trimer. We take the optical
radius of the particle, 517 nm, as an estimate of the particle size and
use the data in Figure \ref{fig:visc} to infer $\eta = 1.049$ mPa\,s,
the value we use in the \textsc{hydrosub} calculations.

\end{document}